\documentclass[letterpaper,twocolumn,english,aps,prl,superscriptaddress]{revtex4-1}

\usepackage[T1]{fontenc}
\usepackage[latin9]{inputenc}
\usepackage{xcolor}
\definecolor{note_fontcolor}{rgb}{0.800781, 0, 0.015625}
\usepackage{babel}
\usepackage{calc}
\usepackage{amsmath}
\usepackage{amssymb}
\usepackage{cancel}
\usepackage{graphicx}
\usepackage{epstopdf}
\usepackage{esint}
\usepackage[unicode=true,pdfusetitle,
 bookmarks=true,bookmarksnumbered=false,bookmarksopen=false,
 breaklinks=true,pdfborder={0 0 0},pdfborderstyle={},backref=false,colorlinks=true]
 {hyperref}
 \usepackage{ulem}
\def\td{\textrm{d}}
\def\i{\textrm{i}}

\def\eq#1{\eqref{#1}}
\def\Eq#1{Eq.~\eqref{#1}}

\def\fig#1{Fig.~\ref{#1}}



\begin{document}
\title{Illustrating the liquid gas transition of nuclear matter in QCD}

\author{Fei Gao}
\affiliation{School of Physics, Beijing Institute of Technology, 100081 Beijing, China}
\author{Yi Lu}
\affiliation{Department of Physics and State Key Laboratory of Nuclear Physics and Technology, Peking University, Beijing 100871, China}

\author{Si-xue Qin}
	\email{sqin@cqu.edu.cn}
	\affiliation{Department of Physics and Chongqing Key Laboratory for Strongly Coupled Physics, Chongqing University, Chongqing 401331, China.}

\author{Zhan Bai}
\email[]{baizhan@siom.ac.cn}
\affiliation{Shanghai Institute of Optics and Fine Mechanics, Chinese Academy of Sciences, Shanghai 201800, China}

 \author{Lei Chang}
\email[ ]{leichang@nankai.edu.cn}
\affiliation{School of Physics, Nankai University, Tianjin 300071, China}
\author{Yu-xin Liu}
\email{yxliu@pku.edu.cn}
\affiliation{Department of Physics and State Key Laboratory of Nuclear Physics and Technology, Peking University, Beijing 100871, China}
\affiliation{Center for High Energy Physics, Peking University, 100871 Beijing, China}
\affiliation{Collaborative Innovation Center of Quantum Matter, Beijing 100871, China}

\date{\today}
\begin{abstract}
We demonstrate that the liquid-gas transition of  nuclear  matter can be rigorously described with the quantum chromodynamics 
by combining the quark gap equation and the Faddeev equation of nucleon.
Our investigation focuses on this transition at zero temperature and finite chemical potential, 
revealing a finite difference between the gas and liquid solution of the quark propagator. 
This difference emerges from the shift of the nucleon pole mass in medium, 
which is generated in the nucleon channel of the quark gap equation.  We prove that such a difference  is precisely the contour contribution  from the shift of the nucleon pole.
The resulting discontinuity manifests as a first-order phase transition and fundamentally determines both the nuclear binding energy and the saturation density.
We then derive an analytical relation between the binding energy and the sigma term of the nucleon,  yielding a binding energy of $E/A=15.9\,\textrm{MeV}$.
Furthermore, by establishing the relation between  the nuclear saturation density and the vector charge of  nucleon in association with the binding energy, 
we determine the saturation density to be  $n_{\textrm{B}}^{0}=0.15\,\textrm{fm}^{-3}$.
\end{abstract}
\maketitle

{\textbf{\textit {Introduction}}. 
The available multi-fragmentation data in nucleus-nucleus and hadron-nucleus reactions demonstrates 
that a critical point exists in nuclear matter for the liquid-gas transition
\citep{Finn:1982tc,Panagiotou:1984rb,Bertsch:1983uv,Bondorf:1995ua,Pochodzalla:1995xy,Richert:2000hp,Chomaz:2003dz,Borderie:2019fii}. 
At zero temperature, the liquid-gas transition is thus characterized as a first-order phase transition.  
The liquid-gas transition of nuclear matter can be phenomenologically described 
by the relativistic mean field model \citep{Walecka:1974qa,Serot:1984ey,Serot:1997xg}.
Within this model, a first-order phase transition can be identified at zero temperature 
and tuned to occur at  baryon chemical potential $\mu^{*}=m_{N}-E/A=923$ MeV, 
with $m_{N}$  the nucleon mass ($m_{N}=939$ MeV) and $E/A=16$ MeV  the binding energy~\cite{Typel:2009sy,Fukushima:2013rx}.  
The Hartree-Fock approach based on the nucleon-nucleon potential 
has also been applied to investigate the liquid-gas transition~\cite{Haftel:1970zz,Baldo:1999cvh,Carbone:2018kji}. Moreover,  the nucleon potential can be calculated directly via the  first principle QCD methods~\cite{Ishii:2006ec,Fukushima:2023wnl}.
However, it remains unclear how to quantitatively describe such a liquid-gas transition 
directly from the fundamental theory of nucleons, namely quantum chromodynamics (QCD).

To fully understand the nuclear matter liquid-gas transition, 
direct investigation of  strong interaction of nucleons in relation to in-medium quark and gluon  is essential.  
One may first consider the ``Silver-Blaze'' property
~\cite{Cohen:2003kd,Cohen:2004qp,Gunkel:2019xnh,Gunkel:2020wcl}, 
indicating that at zero temperature below a certain chemical potential 
the system maintains its vacuum state.  
In Dyson-Schwinger and Faddeev equation frameworks,  
widely used for studying nucleon properties in QCD 
~\cite{Roberts:2007ji,Eichmann:2012zz,Williams:2015cvx,Eichmann:2016yit,Eichmann:2016hgl, Qin:2018dqp},  
this property means the gap and Faddeev equations 
maintain the analytic continuation from vacuum which is valid if  the self-energy kernel has no singularities.

When the kernel of the gap equation encounters singularities, 
the quark propagator deviates from its analytic continuation form, 
causing physical quantities to differ from their vacuum values. 
When this modified quark solution is inserted into the Faddeev equation, the eigenvalue, 
which determines nucleon mass, changes accordingly, 
and the difference between this new mass and the vacuum value yields the nucleon binding energy.  
Thus, solving the gap and Faddeev equations simultaneously 
produces a finite gap between the new solution and the analytic continuation, 
inducing a first-order phase transition that determines both the nucleon binding energy $E/A$ and the saturation density $n^0_B$. 
We therefore propose combining the quark gap equation 
with the nucleon Faddeev equation to elucidate the liquid-gas transition mechanism.

\textbf{\textit{Framework of Dyson-Schwinger Equations}}.
In the Dyson-Schwinger framework,
at zero temperature and finite density, 
the gap equation for quark propagator $S\left(p\right)$ is:
\begin{equation}
\begin{split} 
& S^{-1}\left(p;\mu_q\right)=S_{0}^{-1}\left(\tilde{p}\right)+\Sigma^{\rm G}(p;\mu_q)+\Sigma^{\rm N}(p;\mu_q),\\
\end{split}
\label{eq:DSEq}
\end{equation}
with
\begin{eqnarray}\label{eq:selfenergy} 
&&\Sigma^{\rm G}_{}(p;\mu_{q})=\int\text{d}^{4}{q} D_{\mu\nu}^{ab}\left(p-q\right)\frac{\lambda^{a}}{2}\gamma_{\mu}S\left(q;\mu_q\right)\Gamma_{\nu}^{b}\left(q,p;\mu_q\right) ,\notag\\
&&\Sigma^{\rm N}_{}(p;\mu_{q})=\int\text{d}^{4}{q}_{1}\text{d}^{4}{q}_{2}  S_{}(q_1;\mu_q)\bar{\Gamma}^{(3)}(q_1,q_2,p;\mu_q) \notag\\
&& \times S_{}(q_2;\mu_q)\Gamma^{(3)}(q_1,q_2,p;\mu_q)S_{M_N}\left(\tilde{p}+\tilde{q}_{1}+\tilde{q}_{2}\right) ,\notag
\end{eqnarray}
where $S_{0}\left(p\right)$ is the bare quark propagator  
and $\Sigma^{\rm G, N}$ are self energies shown in Fig.~\ref{fig:gap_eq}. 
In $\Sigma^{\rm G}$, $D_{\mu\nu}^{ab}$ is the gluon propagator and $\Gamma_{\nu}^{b}$ is the quark-gluon vertex. 
$\Sigma^{\rm N}$ involves two quark propagators, one nucleon propagator 
$S_{M_N}\left(P\right)=1/\left(\i\cancel{P}+M_{N}\right)$, 
and the Faddeev amplitude ${\Gamma}^{(3)}$ with its charge conjugation ${\bar\Gamma}^{(3)}$. 
Chemical potential enters through the dressing functions
and via the 4-component of momentum as 
$\tilde{p}=(p_4+\i \mu_q,\vec{p})$, where $\mu_q$ is the quark chemical potential.

\begin{figure}
\includegraphics[width=0.45\textwidth]{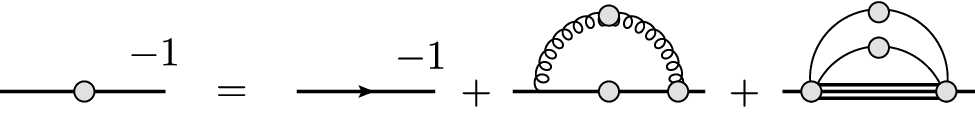}
\caption{The quark gap equation that explicitly includes the nucleon channel with the triple line representing  the nucleon propagator as in Ref.~\cite{Eichmann:2015kfa}. We denote the self energy with nucleon propagator as $\Sigma^{\rm N}$ and the rest of the self energy which excludes the nucleon channel $\Sigma^{\rm N}$ is denoted as $\Sigma^{\rm G}$. \label{fig:gap_eq}}
\end{figure}

\begin{figure}
\includegraphics[width=0.3\textwidth]{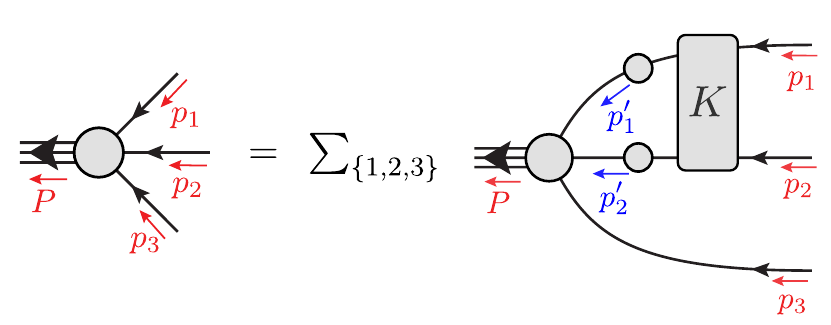}
\caption{Three body Faddeev equation of nucleon in \Eq{eq:Faddeev}. The summation is to sum up the three diagrams for  exchanging the interaction kernel   $K$  between each two quarks. \label{fig:faddeev}}
\end{figure}

For studying the liquid-gas transition, 
we split the original quark gap equation self-energy into a non-nucleon channel $\Sigma^{\rm G}$ and a nucleon channel $\Sigma^{\rm N}$, 
essentially extracted from the full dressed quark-gluon vertex~\cite{Eichmann:2015kfa}. 
$\Sigma^{\rm G}$ resembles the original self-energy kernel but with a different quark-gluon vertex 
since the nucleon channel is now separated. 
The nucleon channel is crucial for the liquid-gas transition, as will be discussed later. 
The Faddeev amplitude in this channel is obtained from the Faddeev equation (Fig.~\ref{fig:faddeev}), 
which is an eigen-equation with $\lambda(P^2)=1$ when $P$ corresponds to the nucleon pole mass. 
This equation reads~\cite{Qin:2018dqp}:
\begin{equation}
\begin{split} 
 \lambda \left(\tilde{P}^2\right)&\Gamma^{(3)}\left(p_1,p_2,P;\mu_q\right)=\int\text{d}^{4}p^\prime_1\text{d}^{4}p^\prime_2\Gamma^{(3)}\left(p^\prime_1,p^\prime_2,P;\mu_q\right)\\
 &\times \left[S\left(p^\prime_{1};\mu_q\right)K\left(p_1,p_2,p^\prime_1,p^\prime_2\right)S\left(p^\prime_{2};\mu_q\right)+\text{cycl.}\right],
\end{split}
\label{eq:Faddeev}
\end{equation}
with $\Gamma^{(3)}$ the Faddeev amplitude of nucleon and $K$ the general interaction kernel as in \fig{fig:faddeev}.

We first describe the ``Silver Blaze property'' at zero temperature 
- the phenomenon where QCD matter observables remain unchanged within a certain chemical potential range. 
This is readily understood from Eqs.~(\ref{eq:DSEq}) and (\ref{eq:Faddeev}). 
For small chemical potential, the integrand contains no singularities, 
allowing the integral variables to be shifted from $\tilde{q},\tilde{p}^\prime$ to $q, p^\prime$ respectively as analytic continuation. 
By replacing the external momentum $\tilde{p}$ with $p$, 
the DSEs at finite chemical potential take the same form as in vacuum. This yields:
\begin{eqnarray}
S(p;\mu_q)=S_{\rm vac}(\tilde{p}),\,\Gamma^{(3)}\left(p_1,p_2,P;\mu_q\right)=\Gamma_{\rm vac}^{(3)}\left(\tilde{p}_1,\tilde{p}_2,\tilde{P}\right).\notag\\
\end{eqnarray}
Here, `$\rm vac$' denotes the corresponding vacuum solution. 
Consequently, the condensate, quark number density, and nucleon properties 
remain unchanged due to the analytic continuation property of Eqs.~(\ref{eq:DSEq}) and (\ref{eq:Faddeev}). 
Thus, within the range $\left[0,\bar\mu_{q}^{}\right)$, 
where $\bar\mu_{q}^{}$ corresponds to the first singularity in the gap equation, 
the system maintains its vacuum properties.

At lowest order, the first singularity emerges from the quark propagator itself, 
with the corresponding chemical potential marking the chiral phase transition. 
However, after incorporating the nucleon channel in higher orders (Fig.~\ref{fig:gap_eq}), 
the nucleon propagator introduces an earlier singularity before the chiral transition, 
corresponding to the liquid-gas transition point. 
Without considering nucleon medium effects, 
one might naively expect this singularity at $\mu_q=M_N/3$.

\begin{figure}
\includegraphics[width=0.3\textwidth]{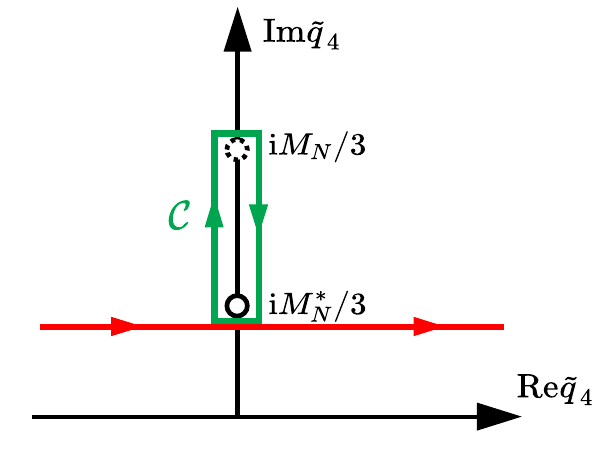}
\caption{
The path of the integral and the singularities in the kernel. 
The red line is the normal  integration path  in the gap equation for the vacuum solution, 
and the green line contour $\mathcal{C}$ is the pole contribution from the nucleon channel in liquid solution, 
which together with the red line builds the complete integration path for the liquid solution. \label{fig:gap_contour}
}
\end{figure}

When nucleon chemical potential exceeds the nucleon mass pole, 
the gap equation's analytical property breaks, 
triggering the liquid-gas transition. 
Simultaneously, the Faddeev equation changes through the quark propagator in its kernel. 
This change shifts the nucleon mass pole, causing the liquid-gas transition point to deviate from the vacuum nucleon mass. 
The difference between vacuum nucleon mass $m_N$ and the mass in the new solution is therefore essential for the first-order phase transition. 
This new ``liquid solution'' contains a new singularity $M_{N}^{\ast}$, 
representing the in-medium nucleon pole mass 
and corresponding to the nuclear liquid-gas phase transition chemical potential at zero temperature. 
Note that $M_{N}^{\ast}$ differs from effective masses in phenomenological models, 
which incorporate average nucleon-nucleon interactions~\cite{Li:2008gp,Li:2018lpy}.

It is important to note that the actual phase transition occurs at $\mu_q=M^*_N/3<M_N/3$ without reaching the singularity at $M_N/3$ as described in the above procedure. Essentially,  the liquid solution represents a self-consistent solution of both the quark gap and the Faddeev equations in medium, 
while such a procedure highlights how the quark propagator's analytic structure connects the  liquid and vacuum solution. 
Following this concept, the quark propagator for the liquid solution can be rewritten as:
\begin{eqnarray}\label{eq:liq}
S_{\text{liq}}^{-1}\left(p;\mu_{q}\right)=  S_{\text{vac},M^*_N}^{-1}\left(\tilde{p}\right)-\delta f_{\mathcal{C}_{}}\left(\tilde{p}\right)S_{\text{vac},M^*_N}^{-1}\left(\tilde{p}\right),
\end{eqnarray}
where $S_{\mathrm{vac},M^*_N}$ is the vacuum solution with nucleon mass replaced by the in-medium mass $M_N^*$, 
and $\delta f_{\mathcal{C}_{}}$ represents the additional difference between the liquid and vacuum solutions beyond the simple nucleon mass replacement. 
This difference, as will be discussed below, arises from the distinct analytic structure of the pole, represented by contour $\mathcal{C}$ in Fig.~\ref{fig:gap_contour}.

The contour contribution is crucial for ensuring the liquid solution truly satisfies the quark gap equation. 
Merely replacing the nucleon mass modifies the nucleon channel without feedback. It thus 
produces an incomplete solution to the gap equation,  and  the consequential change of the quark propagator  must also be incorporated into the self-energy calculation. 
This additional modification corresponds precisely to contour $\mathcal{C}$'s contribution, 
since the contour contribution completes the change in the quark propagator  from simple mass replacement to proper analytic continuation 
with the singularity shifting from $M_N/3$ in vacuum to $M^*_N/3$ in medium.

Therefore, 
$\delta f_{\mathcal{C}_{}}$ can be defined as the nuclear propagator's pole contribution in Fig.~\ref{fig:gap_eq} from contour $\mathcal{C}$  as:
\begin{eqnarray}\label{eq:contour} 
&& \delta f_{\mathcal{C}_{}}\left(\tilde{p}\right)S^{-1}_{\text{vac},M^*_N}\left(\tilde{p}\right)=\varoiint^{\rm N}_{\mathcal C}\text{d}^{4}q_{1}\text{d}^{4}q_{2}\\
&&\left[S_{\text{vac},M^*_N}\left(\tilde{q}_{1}\right)\bar{\Gamma}^{(3)}_{\rm vac}S_{\text{vac},M^*_N}\left(\tilde{q}_{2}\right)\Gamma^{(3)}_{\rm vac}S_{\text{N}}\left(\tilde{p}-\tilde{q}_{1}-\tilde{q}_{2}\right)\right],\notag
\end{eqnarray}
where the contour integral captures only the nucleon propagator pole contribution, 
from $M_N^*/3$ to $\mu_q+(M_N-M^*_N)/3$ as shown in Fig.~\ref{fig:gap_contour}. 
This liquid solution construction satisfies the quark gap equation by ensuring the quark propagator maintains analytic continuation in the self-energy, 
with  the only change in self-energy being the explicit term from $\delta f_{\mathcal{C}_{}}\left(\tilde{p}\right)$. 
A detailed proof appears in the supplemental material.

\textbf{\textit{Liquid Gas Transition and  Nucleon Charge.}}
From above analysis, 
to calculate the nuclear liquid-gas phase transition, 
we must simultaneously solve the gap equation with baryon back-reaction and the Faddeev equation. 
Rather than solving these coupled equations numerically, in the present work we apply a simple approximation to estimate the difference between vacuum and liquid solutions.

We now approximate $\delta f_{\mathcal{C}_{}}(\tilde{p})$ as a constant  at its infrared limit $\delta\bar{f}=\frac{1}{4}\textrm{Tr}[\delta f_{\mathcal{C}_{}}(p=0)]$  since  the liquid gas transition is related to the nucleon property which is dominant by the  infrared property of quark. One  has:
\begin{equation}
S_{\text{liq}}^{-1}\left(p;\mu_{q}\right)=(1-\delta \bar f)S_{\text{vac},M^*_N}^{-1}\left(\tilde p\right).
\end{equation}
%Inserting this liquid solution  into the Faddeev equation induces an eigenvalue correction. 

Under this approximation,  
we obtain a simple relation between the modified eigenvalue at the pole position and the original vacuum eigenvalue through inserting the liquid solution into the Faddeev equation in \Eq{eq:Faddeev} that reads:
\begin{eqnarray}\label{eq:Inmed}
\lambda(p^{2})=\frac{\lambda^{\text{vac}}(p^{2})}{\left(1-\delta\bar{f}\right)^{2}}\approx\frac{\lambda^{\text{vac}}(p^{2})}{1-2\delta\bar{f}}.
\end{eqnarray}
The power of two arises because the dominant contribution of the Faddeev equation kernel $K$ is one-gluon exchange between two quark propagators.
This relation indicates that the Faddeev equation's eigenvalue at the liquid-gas transition point differs from its vacuum counterpart, 
as does the effective mass $m_{N}^{\ast}$. 
Recalling the Nakanishi normalization condition for bound states~\citep{Nakanishi:1965zza,Yao:2024uej}:
\begin{eqnarray}
\frac{\partial\log(\lambda^\text{vac}(p^{2}))}{\partial \log (-p^{2}/M^2_N)}\Bigg|_{p^{2}=-M_N^{2}}=Z,\label{eq:nakanishi}
\end{eqnarray}
where $Z$ is the Nakanishi normalization factor computed numerically by solving the Faddeev equation. 
For nucleons, one has $Z=0.97$~\citep{Yao:2024uej}.
%\begin{eqnarray}
%Z=0.97.
%\end{eqnarray}

Since the difference between $M_N$ and $M^*_N$ is small, we can expand the eigenvalue in Eq.~(\ref{eq:nakanishi}) at the vacuum pole mass:
\begin{eqnarray}\label{eq:lambdavac}
\lambda^{\text{\text{vac}}}(p^{2}=-M_{N}^{*2})=1+Z\frac{M_{N}^{*2}-M^2_{N}}{M^2_{N}},
\end{eqnarray}
With the in-medium nucleon eigenvalue $\lambda(p^{2}=-M_{N}^{*2})=1$ together with Eq.~(\ref{eq:Inmed}) and (\ref{eq:lambdavac}), we obtain:
\begin{eqnarray}
M_{N}^{*}=(1-\delta\bar{f}/Z)M_{N}.
\end{eqnarray}

Inserting this relation into the gap equation, 
one could directly solve the gap equation under certain truncations as in Ref.~\cite{Eichmann:2015kfa} to obtain $\delta\bar{f}$. 
Instead, we apply a model-independent analysis by relating liquid-gas transition properties to nucleon charges. 
First, the factor $\delta\bar{f}$ can be estimated via the nucleon scalar charge related to the scalar vacuum bubble. 
As shown in Fig.~\ref{fig:vacuum} and defined in Ref.\cite{Eichmann:2015kfa}, the scalar vacuum bubble $\mathcal S$ is written as:
\begin{eqnarray}
\mathcal{S}=\frac{2M_{N}}{3}\int \frac{d^{4}q}{(2\pi)^4}\frac{g_{s}(q^2,0)}{{q^{2}+M_{N}^{2}}}.
\end{eqnarray}

\begin{figure}
\includegraphics[width=0.15\textwidth]{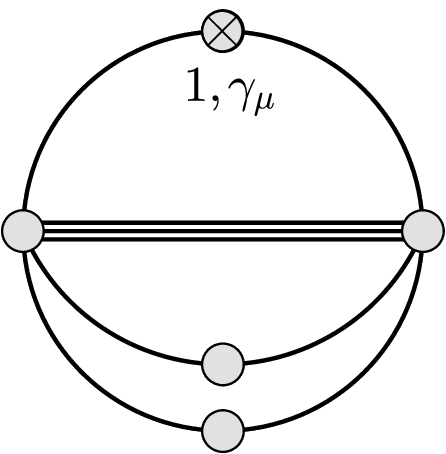}
\caption{The  vacuum bubble with scalar and the vector vertex, which  become the scalar and vector charge of nucleon respectively after putting the nucleon propagator on shell. \label{fig:vacuum}}
\end{figure}

To match with the definition of $\delta f_{\mathcal C}$ in gap equation,  here we only account the   pole contribution from nucleon propagator in the  vacuum bubble  of liquid solution
in  the contour integral following contour $\mathcal{C}$  as in \fig{fig:gap_contour} , which then  reads:
\begin{equation}
\begin{split}
&\delta_{\mathcal{C}}\mathcal{S}=\frac{2M^*_{N}}{3} \oint_{\mathcal C} \frac{d q_4}{2\pi}\int\frac{d^{3}q}{(2\pi)^3}\frac{g_{s}(-M_{N}^{2},0)}{{q^{2}+(M^*_{N})^{2}}}\\
=&\frac{2M^*_{N}}{3} \int \frac{d^{3}q}{(2\pi)^3}\theta(|\vec{q}|-\sqrt{M_{N}^{2}-(M_{N}^{*})^{2}})\frac{g_{s}(-M_{N}^{2},0)}{2\sqrt{q^{2}+(M_{N}^*)^{2}}}.\label{eq:delta_S1}
\end{split}
\end{equation}
where scalar charge has been put on shell  as $g_{s}(q^2=-M_{N}^{2},0)=\sigma_{N}/(2m_{q})$ since $\delta \bar f$ is small and hence contour $\mathcal{C}$ is always in the neighborhood of $q^2=-M_{N}^{2}$, and  $\sigma_N$  is  the nucleon sigma term and $m_q$ the current quark mass. 
The vacuum bubble can be related to the factor $\delta \bar{f}$ with Eq.~(\ref{eq:contour}). 
By attaching quark propagator legs to respective scalar and vector vertices, 
the nucleon channel of the quark  gap equation in Fig.~\ref{fig:gap_eq} transforms into the vacuum bubble in Fig.~\ref{fig:vacuum}, yielding:
\begin{equation}
\begin{split}
\delta_{\mathcal C}\mathcal{S}=&\frac{1}{2}\oint_{\mathcal C}\frac{d^{4}q}{(2\pi)^4} \rm{Tr}[ S_{\rm vac,M_N^*}\Sigma^{\rm N}(q;\mu_q)S_{\rm vac,M_N^*}]\\
=&\frac{1}{2}\int \frac{d^{4}q}{(2\pi)^4}\textrm{Tr}[S_{\rm vac,M_N^*}^{}\delta f_{\mathcal{C}}(q;\mu_q)]\cong\frac{1}{2}\langle\bar{q}q\rangle\delta\bar{f},\label{eq:delta_S2}
\end{split}
\end{equation}
where $\langle\bar{q}q\rangle$ is the chiral condensate. 
Since $\delta \bar{f}$ is small in Eq.~(\ref{eq:delta_S2}), 
combining Eqs.~(\ref{eq:delta_S1}) and (\ref{eq:delta_S2}) yields an analytic form:
\begin{equation}
\delta \bar{f}=\frac{81\pi^4Z^3M_\pi^4f_\pi^4}{8M_N^6\sigma_N^2}=0.0164,\label{eq:df}
\end{equation}
with the GMOR relation $2m_q\langle\bar{q}q\rangle=M_\pi^2f_\pi^2$ with $M_\pi=135$ MeV and $f_\pi=92$ MeV. 
The sigma term is taken as $\sigma_N=43.7$ MeV from recent lattice QCD simulations~\cite{Agadjanov:2023efe}. 
This gives $M_{N}^{\text{\ensuremath{\ast}}}=939-15.9$ MeV, 
and  at $\mu_{B}=M_{N}^{*}=923.1$ MeV, 
the system transits  from vacuum to liquid solution with a finite gap, resulting in a first-order liquid-gas transition.
This result agrees well with the typical value of $\mu_{B}=923\;$MeV.

Moreover,  one can similarly relate the number density to the nucleon vector charge at the liquid-gas transition point. 
Analogous to the relationship between scalar vacuum bubble and scalar charge, 
the vector vacuum bubble difference is defined as:
\begin{equation}
\begin{split}
\delta_{\mathcal C}\mathcal{V}=\frac{2M^*_{N}}{3}\int& \frac{d^{3}q}{(2\pi)^3}\theta(|\vec{q}|-\sqrt{M_{N}^{2}-(M_{N}^{*})^{2}})\\
&\times\frac{g_{v}(-M_{N}^{2},0)}{2\sqrt{q^{2}+(M_{N}^*)^{2}}},\label{eq:VaccumBubble1}
\end{split}
\end{equation}
where the vector charge $g_{v}(-M_{N}^{2},0)=1$. 
This vacuum bubble relates to the factor $\delta f(p^{2})$ and quark number density $\langle\bar{q}\gamma_0 q\rangle$ as:
\begin{eqnarray}\label{eq:VaccumBuuble2}
\delta_{\mathcal C}\mathcal{V}&=&\frac{1}{2}\oint_{\mathcal C}\frac{d^{4}q}{(2\pi)^4} \rm{Tr}[\gamma_4 S_{\rm vac,M_N^*}\Sigma^{\rm N}(q;\mu_q)S_{\rm vac,M_N^*}]
\\
&=&\frac{1}{2}\int \frac{d^{4}q}{(2\pi)^4}\textrm{Tr}[\gamma_4S_{\rm vac,M_N^*}^{}\delta f_\mathcal{C}(q^{2})]\cong\frac{1}{2}  \langle\bar{q}\gamma_4 q\rangle\delta\bar{f}.\notag
\end{eqnarray}

The saturation density of nucleon at liquid gas phase transition point is then given as:
\begin{equation}
n_{\rm B}^0=\frac{1}{3}\langle\bar{q}\gamma_4 q\rangle=\frac{\sqrt{8\delta f/Z^3}}{27\pi^2}M^3_N=0.15 \,\textrm{fm}^{-3}.
\end{equation}

One can also insert \Eq{eq:df} in definition of the saturation density which then yields a simple relation as:
\begin{equation}
n_{\rm B}^0=\frac{M_\pi^2f_\pi^2}{3\sigma_N}.
\end{equation}
The relation is valid for the chemical potential near above the liquid gas phase transition point where the approximation described above is valid.

\textbf{\textit{Summary}}.
Within the Dyson-Schwinger and Faddeev equation framework, 
we establish a method to describe nuclear liquid-gas phase transition directly from quantum chromodynamics. 
We elucidate the transition mechanism through the quark propagator's analytical structure, 
particularly  the singularity from the nucleon propagator in the nucleon channel of the quark gap equation. 
Before this singularity, the gap equation satisfies analytic continuation, manifesting the ``Silver-Blaze'' property. 
When chemical potential increases and the quark propagator integrand encounters the singularity, 
the solution deviates from its analytic continuation form. 
Simultaneously, the Faddeev equation changes because its kernel contains quark propagators. 
This dual change in quark propagator and nucleon Faddeev amplitude creates a finite pole position shift between liquid and vacuum solutions, 
which ultimately determines the nucleon binding energy and induces a first-order transition at the new solution's pole mass. The constructed liquid solution can be proved to satisfy the quark gap equation.

Given the small difference between liquid and vacuum solutions, 
we derive analytic expressions for the binding energy $E/A$ and the saturation density $n_0$ by relating them to nucleon scalar and vector charges respectively.
We obtain $E/A=15.9$ MeV, in excellent agreement with experimental data. 
Similarly calculated saturation density yields $n_B^0=0.15$ fm$^{-3}$, 
consistent with expected values. 
Notably, we discover a simple relation $n_B^0=M_\pi^2 f_\pi^2/(3\sigma_N)$ connecting saturation density to the nucleon sigma term, 
which can constrain phenomenological parameters in low-energy QCD effective models.

%Considering the difference between the liquid solution and the vacuum solution being small, it is possible to give the analytic expressions for the binding energy $E/A$ and the saturation density $n_0$ at phase transition point after building its relation to the scalar and vector charge of nucleon respectively. Specifically, only taking into account the contour contribution of the singularity from such a difference,  we obtain the  binding energy  as   $E/A=15.9$ MeV which is in very good agreement  with the experimental data. The saturation density of nucleon can be also computed similarly and we get $n_0=0.15$ fm$^{-3}$ which is consistent with  the expected value.  Moreover, we find a simple relation between $n_0$ and the nucleon Sigma term as $n_0=M_\pi^2 f_\pi^2/(3\sigma_N)$. This can help to  constrain the phenomenological parameters in the low energy effective models of QCD.  

%\begin{section}{Acknowledgements}
\textbf{\textit{Acknowledgements}}.
FG and YL thank  Jan M. Pawlowski, Weijie Fu, Chuang Huang %and other members of the fQCD collaboration
for discussions.
ZB is supported by National Natural Science Foundation of China (No. 12388102), 
the Strategic Priority Research Program of the Chinese Academy of Sciences (No. XDB0890303),
and the CAS Project for Young Scientists in Basic Research (YSBR060).
FG is  supported by the National  Science Foundation of China under Grants  No. 12305134. YL and YXL are supported by the National Natural Science Foundation of China under Grants  No. 12247107 and  No. 12175007.
LC is  supported by the National  Science Foundation of China under  Grant no. 12135007.
%\end{section}

\bibliographystyle{apsrev4-1}
\bibliography{Ref-LiquidGas}

\onecolumngrid
\appendix
\newpage{}

\section*{Supplemental Material: the proof for the validity of the construction of the liquid solution} 
\label{App}
First of all, one defines a new solution that is based on the vacuum
solution but with a shift of nucleon mass $M_{N}\rightarrow M_{N}^{\ast}$
in the gap equation, i.e., we consider the function $S_{\text{vac},M_{N}^{\ast}}$
which is the solution of the following equation: 
\begin{equation}
\begin{split}S_{\text{vac},M_{N}^{*}}^{-1}\left(\tilde{p}\right)= & S_{0}^{-1}\left(\tilde{p}\right)+\int\td^{4}qD_{\mu\nu}^{ab}\left(\tilde{p}-\tilde{q}\right)\frac{\lambda^{a}}{2}\gamma_{\mu}S_{\text{vac},M_{N}^{*}}\left(\tilde{q}\right)\Gamma_{\nu}^{b}\left(\tilde{q},\tilde{p}\right)\\
+\int\td^{4}q_{1}\int\td^{4}q_{2} & S_{\text{vac},M_{N}^{*}}\left(\tilde{q}_{1}\right)\bar{\Gamma}_{\text{vac}}^{\left(3\right)}\left(\tilde{q}_{1},\tilde{q}_{2},\tilde{p}\right)S_{\text{vac},M_{N}^{*}}\left(\tilde{q}_{2}\right)\Gamma_{\text{vac}}^{\left(3\right)}\left(\tilde{q}_{1},\tilde{q}_{2},\tilde{p}\right)S_{M_{N}^{\ast}}\left(\tilde{p}+\tilde{q}_{1}+\tilde{q}_{2}\right),
\end{split}
\label{eq:gape_Svac*}
\end{equation}
with $\tilde{p}=\left(p_{4}+\i\mu_{q}^{\ast},\vec{p}\right)$ and
$\mu_{q}^{*}=M_{N}^{*}/3-0^{+}$. 
Now considering the integral deformation of the self-energy from $\tilde{q}$
(similarly for $\tilde{q}_{1}$ and $\tilde{q}_{2}$) to $\bar{q}$
with $\bar{q}=\left(q_{4}+\i\bar{\mu}_{q},\vec{q}\right)$ and $\bar{\mu}_{q}=\mu_{q}^{*}+\frac{M_{N}-M_{N}^{*}}{3}=M_{N}/3$,
one  gets immediately for the self energy $\Sigma^{\text{G}}$: 
\begin{equation}
\begin{split} & \int\td^{4}qD_{\mu\nu}^{ab}\left(\tilde{p}-\bar{q}\right)\frac{\lambda^{a}}{2}\gamma_{\mu}S_{\text{vac},M_{N}^{*}}\left(\bar{q}\right)\Gamma_{\nu}^{b}\left(\bar{q},\tilde{p}\right)\\
= & \int\td^{4}qD_{\mu\nu}^{ab}\left(\tilde{p}-\tilde{q}\right)\frac{\lambda^{a}}{2}\gamma_{\mu}S_{\text{vac},M_{N}^{*}}\left(\tilde q\right)\Gamma_{\nu}^{b}\left(\tilde{q},\tilde{p}\right)+\int\td^{3}\vec{q}\oint_{\mathcal{C}}\td\tilde{q}_{4}D_{\mu\nu}^{ab}\left(\tilde{p}-\tilde{q}\right)\frac{\lambda^{a}}{2}\gamma_{\mu}S_{\text{vac},M_{N}^{*}}\left(\tilde q\right)\Gamma_{\nu}^{b}\left(\tilde{q},\tilde{p}\right),
\end{split}
\label{eq:G_continue}
\end{equation}
with $\mathcal{C}$ the contour defined in \fig{fig:gap_contour}.
For short notation, we define the last line in Eq.~(\ref{eq:G_continue})
as $\Sigma_{\mathcal{C}}^{\text{G}}$.

For the nucleon self energy $\Sigma^{\text{N}}$, we have: 
\begin{equation}
\begin{split} & \int\td^{4}q_{1}\td^{4}q_{2}S_{\text{vac},M_{N}^{*}}\left(\bar{q}_{1}\right)\bar{\Gamma}_{\text{vac}}^{\left(3\right)}\left(\bar{q}_{1},\bar{q}_{2},\tilde{p}\right)S_{\text{vac},M_{N}^{*}}\left(\bar{q}_{2}\right)\Gamma_{\text{vac}}^{\left(3\right)}\left(\bar{q}_{1},\bar{q}_{2},\tilde{p}\right)S_{M_{N}^{\ast}}\left(\tilde{p}+\bar{q}_{1}+\bar{q}_{2}\right)\\
= & \int\td^{4}q_{1}\td^{4}q_{2}S_{\text{vac},M_{N}^{*}}\left(\tilde{q}_{1}\right)\bar{\Gamma}_{\text{vac}}^{\left(3\right)}\left(\tilde{q}_{1},\tilde{q}_{2},\tilde{p}\right)S_{\text{vac},M_{N}^{*}}\left(\tilde{q}_{2}\right)\Gamma_{\text{vac}}^{\left(3\right)}\left(\tilde{q}_{1},\tilde{q}_{2},\tilde{p}\right)S_{M_{N}^{\ast}}\left(\tilde{p}+\tilde{q}_{1}+\tilde{q}_{2}\right)\\
+ & \left[\oiint_{\mathcal{C}}^{S}+\oiint_{\mathcal{C}}^{N}\right]\td^{4}\tilde{q}_{1}\td^{4}\tilde{q}_{2}S_{\text{vac},M_{N}^{*}}\left(\tilde{q}_{1}\right)\bar{\Gamma}_{\text{vac}}^{\left(3\right)}\left(\tilde{q}_{1},\tilde{q}_{2},\tilde{p}\right)S_{\text{vac},M_{N}^{*}}\left(\tilde{q}_{2}\right)\Gamma_{\text{vac}}^{\left(3\right)}\left(\tilde{q}_{1},\tilde{q}_{2},\tilde{p}\right)S_{M_{N}^{\ast}}\left(\tilde{p}+\tilde{q}_{1}+\tilde{q}_{2}\right),
\end{split}
\label{eq:N_continue}
\end{equation}
where $\oiint_{\mathcal{C}}^{S}$ is the pole contribution from the
quark propagator in the contour $\mathcal{C}$ denoting as $\Sigma_{\mathcal{C}}^{\text{N}}$,
while $\oiint_{\mathcal{C}}^{N}$ is the pole contribution from the
nucleon propagator and has been denoted as $\delta f_{\mathcal{C}}\left(\tilde{p}_{}\right)S_{\text{vac},M_{N}^{*}}^{-1}$
as in \Eq{eq:contour}.

One firstly consider the leading order of $\delta f_{\mathcal{C}}$
in the self-energy. The leading order term is not only the explicit
term $\delta f_{\mathcal{C}}$, but also included in $\Sigma_{\mathcal{C}}^{\text{G}}$
and $\Sigma_{\mathcal{C}}^{\text{N}}$, as the solution $S_{\text{vac},M_{N}^{*}}$
satisfies the gap equation which again includes $\delta f_{\mathcal{C}}$.
Inserting Eq.~(\ref{eq:gape_Svac*}) into $\Sigma_{\mathcal{C}}^{\text{G,N}}$,
one has: 
\begin{equation}
\begin{split}\label{eq:contourG}
 & \Sigma^{G}_{\mathcal 
C}=\int\td^{3}\vec{q}\oint_{\mathcal{C}}\td{q}_{4}D_{\mu\nu}^{ab}\left(\tilde{p}-\tilde{q}\right)\frac{\lambda^{a}}{2}\gamma_{\mu}S_{\text{vac}, M^*_N}\left(\tilde q\right)\Sigma^N S_{\text{vac}, M^*_N}\left(\tilde q\right)\Gamma_{\nu}^{b}\left(\tilde{q},\tilde{p}\right)+\hat{\mathcal O}(\delta f_{\mathcal C}^2)\\
 &=\int\td^{4}{q}D_{\mu\nu}^{ab}\left(\tilde{p}-\tilde{q}\right)\frac{\lambda^{a}}{2}\gamma_{\mu}S_{\text{vac}, M^*_N}\left(\tilde q\right)\delta f_{\mathcal C}\left(\tilde{q}\right)\Gamma_{\nu}^{b}\left(\tilde{q},\tilde{p}\right)+\hat{\mathcal O}(\delta f_{\mathcal C}^2), 
\end{split}
\end{equation}
and similarly, 
\begin{equation}
\begin{split}\Sigma_{\mathcal{C}}^{\text{N}}= & 2\int\td^{4}\tilde{q}_{1}\td^{4}\tilde{q}_{2}S_{\text{vac},M_{N}^{*}}\left(\tilde{q}_{1}\right)\bar{\Gamma}_{\text{vac}}^{\left(3\right)}\left(\tilde{q}_{1},\tilde{q}_{2},\tilde{p}\right)S_{\text{vac},M_{N}^{*}}\left(\tilde{q}_{2}\right)\delta f_{\mathcal{C}}\left(\tilde{q}_{1,2}\right)\\
 & \times\Gamma_{\text{vac}}^{\left(3\right)}\left(\tilde{q}_{1},\tilde{q}_{2},\tilde{p}\right)S_{M_{N}^{\ast}}\left(\tilde{p}+\tilde{q}_{1}+\tilde{q}_{2}\right)+\hat{\mathcal{O}}(\delta f_{\mathcal{C}}^{2}),
\end{split}
\label{eq:contourN}
\end{equation}
With the above observations, we then construct the liquid solution
as in \Eq{eq:liq} as: 
\begin{eqnarray*}
S_{\text{liq}}^{-1}\left(p;\mu_{q}^{*}\right)=S_{\text{vac},M_{N}^{*}}^{-1}\left(\tilde{p}\right)-\delta f_{\mathcal{C}}\left(\tilde{p}\right)S_{\text{vac},M_{N}^{*}}^{-1}\left(\tilde{p}\right).
\end{eqnarray*}
Since $\delta f_{\mathcal{C}}$ is expected as small perturbation,
we can expand $S_{\text{liq}}$ and keep only the leading order  as: 
\begin{equation}
S_{\text{liq}}=S_{\text{vac},M_{N}^{*}}\left(1-\delta f_{\mathcal{C}}\right)^{-1}\approx S_{\text{vac},M_{N}^{*}}\left(1+\delta f_{\mathcal{C}}\right).\label{eq:liq_expand}
\end{equation}

The $\delta f_{\mathcal{C}}$ term involved in the construction is
to cancel the contribution of $\Sigma_{\mathcal{C}}^{{\rm G,N}}$
as will be illustrated as follows. We intend to verify that the liquid solution satisfies the gap equation,
i.e., we want to prove the following equation 
\begin{equation}
\begin{split}S_{\text{liq}}^{-1}\left(p;\mu_{q}^{\ast}\right)\overset{?}{=} & S_{0}^{-1}\left(\tilde{p}\right)+\int\td^{4}qD_{\mu\nu}\left(\tilde{p}-\tilde{q}\right)\frac{\lambda^{a}}{2}\gamma_{\mu}S_{\text{liq}}\left(q;\mu_{q}^{\ast}\right)\Gamma_{\nu}^{b}\left(\tilde{q},\tilde{p}\right)\\
 & +\int\td^{4}q_{1}\int\td^{4}q_{2}S_{\text{liq}}\left(q_{1};\mu_{q}^{\ast}\right)\bar{\Gamma}_{\text{liq}}^{\left(3\right)}\left(\tilde{q}_{1},\tilde{q}_{2},\tilde{p}\right)S_{\text{liq}}\left(q_{2};\mu_{q}^{\ast}\right)\Gamma_{\text{liq}}^{\left(3\right)}\left(\tilde{q}_{1},\tilde{q}_{2},\tilde{p}\right)S_{M_{N}^{\ast}}\left(\tilde{p}+\tilde{q}_{1}+\tilde{q}_{2}\right).
\end{split}
\label{eq:gape_Sliq?}
\end{equation}

Substituting Eq.~(\ref{eq:liq_expand})  into the RHS
of Eq.~(\ref{eq:gape_Sliq?}), we have: 
\begin{align*}
 & \text{RHS}=S_{0}^{-1}\left(\tilde{p}\right)\\
& +\left\{ \int\td^{4}qD_{\mu\nu}^{ab}\left(\tilde{p}-\tilde{q}\right)\frac{\lambda^{a}}{2}\gamma_{\mu}S_{\text{vac},M_{N}^{\ast}}\left(\tilde{q}\right)\Gamma_{\nu}^{b}\left(\tilde{q},\tilde{p}\right)+\int\td^{4}qD_{\mu\nu}^{ab}\left(\tilde{p}-\tilde{q}\right)\frac{\lambda^{a}}{2}\gamma_{\mu}S_{\text{vac},M_{N}^{\ast}}\left(\tilde{q}\right)\delta f_{\mathcal{C}}\left(\tilde{q}_{}\right)\Gamma_{\nu}^{b}\left(\tilde{q},\tilde{p}\right)\right\} \\
 & +\left\{ \int\td^{4}q_{1}\int\td^{4}q_{2}S_{\text{vac},M_{N}^{\ast}}\left(\tilde{q}_{1}\right)\bar{\Gamma}_{\text{vac}}^{\left(3\right)}\left(\tilde{q}_{1},\tilde{q}_{2},\tilde{p}\right)S_{\text{vac},M_{N}^{\ast}}\left(\tilde{q}_{2}\right)\Gamma_{\text{vac}}^{\left(3\right)}\left(\tilde{q}_{1},\tilde{q}_{2},\tilde{p}\right)S_{M_{N}^{\ast}}\left(\tilde{p}+\tilde{q}_{1}+\tilde{q}_{2}\right)\right.\\
 & +\left.2\int\td^{4}\tilde{q}_{1}\td^{4}\tilde{q}_{2}S_{\text{vac},M_{N}^{\ast}}\left(\tilde{q}_{1}\right)\bar{\Gamma}_{\text{vac}}^{\left(3\right)}\left(\tilde{q}_{1},\tilde{q}_{2},\tilde{p}\right)S_{\text{vac},M_{N}^{\ast}}\left(\tilde{q}_{2}\right)\delta f_{\mathcal{C}}\left(\tilde{q}_{1,2}\right)\left(\tilde{q}\right)\,\Gamma_{\text{vac}}^{\left(3\right)}\left(\tilde{q}_{1},\tilde{q}_{2},\tilde{p}\right)S_{M_{N}^{\ast}}\left(\tilde{p}+\tilde{q}_{1}+\tilde{q}_{2}\right)+\delta f_{\mathcal{C}}(\tilde p)S_{\text{vac},M_{N}^{\ast}}^{-1}\right\} \\
 & -\delta f_{\mathcal{C}}(\tilde p)S_{\text{vac},M_{N}^{\ast}}^{-1}
\end{align*}
where the difference between the $\Gamma_{\text{liq}}^{(3)}$ and
$\Gamma_{\text{vac}}^{(3)}$ has been neglected as it is the higher order of $\delta f_{\mathcal C}$.

One may  notice immediately that the formula inside $\{...\}$ is precisely
 the integral deformation of  $\Sigma^{\text{G}}$ and $\Sigma^{\text{N}}$
of the vacuum solution $S_{\text{vac},M_{N}^{*}}$, see Eqs.(\ref{eq:G_continue})
and (\ref{eq:N_continue}) together with Eqs.\eq{eq:contourG} and \eq{eq:contourN} . Therefore, one has: 
\begin{align*}
\text{RHS}= & S_{0}^{-1}\left(\tilde{p}\right)+\Sigma_{\text{liq}}^{\text{G}}+\Sigma_{\text{liq}}^{\text{N}}\\
= & S_{0}^{-1}+\Sigma_{\text{vac},M_{N}^{\ast}}^{\text{G}}+\Sigma_{\text{vac},M_{N}^{\ast}}^{\text{N}}-\delta f_{\mathcal{C}}S_{\text{vac,}M_{N}^{\ast}}^{-1}\\
= & S_{\text{vac},M_{N}^{\ast}}^{-1}-\delta f_{\mathcal{C}}S_{\text{vac},M_{N}^{\ast}}^{-1}\\
\equiv & S_{\text{liq}}^{-1}=\text{LHS},
\end{align*}
i.e., $S_{\text{liq}}$ satisfies the gap equation.

\end{document}